\documentclass[12pt]{article}
\usepackage{amsmath,amssymb}
\renewcommand{\a}{\alpha}
\renewcommand{\b}{\beta}
\newcommand{\bs}{\bigskip}
\newcommand{\dt}{\cdot}

\newcommand{\g}{\gamma}

\renewcommand{\i}{\infty}
\renewcommand{\l}{\lambda}
\renewcommand{\L}{\Lambda}
\newcommand{\mb}{\mbox}

\renewcommand{\th}{\theta}

\begin{document}
\setlength{\baselineskip}{18pt}

\begin{center}
{\large\bf A Note Concerning the}\smallskip\\
{\large\bf Turbulent Boundary Layer Drag at Large Reynolds Numbers}

\vspace{1 truein}

{\sc G. I. Barenblatt,${}^1$ \ A. J. Chorin${}^1$ and V. M.
Prostokishin${}^2$}

\vspace{.33 truein}
${}^1$Department of Mathematics and
Lawrence Berkeley National Laboratory,\\
University of California,
Berkeley, California 94720--3840;\\
${}^2$P. P. Shirshov Institute of Oceanology, Russian Academy of
Science,\\ 36 Nakhimov Prospect, Moscow 117218 Russia
\end{center}

\bs\bs
\begin{quote}
{\bf Abstract.} A correlation is obtained  for the drag coefficient
$c\,'_f$ of the turbulent boundary layer as a function of the effective
boundary layer Reynolds number $Re$ that we previously introduced. A
comparison is performed also with another correlation for the drag  coefficient
  as a function of the traditional Reynolds number
$Re_{\th}$, based on the momentum thickness of the boundary layer
proposed recently by R.~D.~Watson, R.~M.~Hall and J.~B.~Anders ({\sc nasa}
Langley Research Center) on the basis of different set of experimental data.
We show that the correlation obtained by us agrees with experimental 
data from the
Illinois Institute of Technology, but is incompatible with the data 
obtained in the
Royal Institute of Technology at Stockholm. On the other hand, both 
sets of data are
in disagreement with the Langley correlation.
\end{quote}

\newpage
\section{Introduction}

The drag coefficient for the turbulent boundary layer is determined as follows:
$$
c\,'_f = \frac{2u_*^2}{U^2} \ , \eqno [1]
$$
where $u_*$ is the friction velocity,
$u_*\!=\!\sqrt{\tau/\rho}$,  $\tau$ the wall shear
stress, $\rho$ the fluid density, $U$ is the free stream velocity.
It is apparently the most practically important  characteristic of 
the turbulent
boundary layer. In the present note a correlation  will be derived 
between the drag
coefficient $c\,'_f$ and the {\it effective} Reynolds number
$Re$, introduced previously (1,2) and determined as follows. It was 
shown (2) that
when the free stream turbulence is low, the average velocity $u$ 
distribution in the
intermediate region of the boundary layer between the viscous 
sublayer and the free
stream consists of two different self-similar structures described by 
the scaling
(power) laws $\phi=A\eta^{\a}$ and $\phi=B\eta^{\b}$. Here $\phi=u/u_*$ and
$\eta=u_*y/\nu$ are  dimensionless variables (recently the notations
$U^+$ and $y^+$ instead of classical notations $\phi$ and $\eta$ have 
become more
popular),
$\nu$ the fluid kinematic viscosity,
$y$ the distance from the boundary. The interface between the two 
regions described
by different scaling laws is sharp at $y=\l$, so that the viscous 
sublayer and the
first region adjacent to it form the sharply bounded wall layer of 
thickness~$\l$.
Furthermore, we showed (2) that at sufficiently large  Reynolds 
numbers the mean
velocity distribution in the first region can be represented in the 
form obtained
earlier for pipes (see~(3))
$$
\phi =\frac{u}{u_*} =\left(\frac{1}{\sqrt{3}} \ln Re +\frac 52\right)
\left(\frac{u_*y}{\nu}\right)^{\frac{3}{2\ln Re}} \eqno [2]
$$
where $Re=U\Lambda/\nu$ is a  Reynolds number, defined by a
characteristic length scale $\L$  proportional to the thickness $\l$ 
of the wall
layer (4). The procedure for determining the effective Reynolds number and the
characteristic length $\L$, described in detail in (1,2), is not more
complicated, perhaps even simpler than the procedure for determining 
the momentum
thickness $\th$ (1,4).

If the effective Reynolds number $Re$ is in fact a governing 
characteristic of the
turbulent boundary layer, like the Reynolds number based on the 
average velocity and
pipe diameter for  flow in pipes, then the drag coefficient $c\,'_f$ 
should be a
function of this parameter. Therefore it was natural to correlate the 
experimental
data for drag coefficient and the effective Reynolds number 
determined for the same
experiments. Such correlation will be presented and discussed below.

On the other hand, Watson, Hall and Anders (5) at the {\sc nasa}
Langley Research Center recently obtained a different correlation of the drag
coefficient with the parameter traditional in turbulent boundary layer studies
  --- the Reynolds number $Re_{\th}$ based on the momentum thickness. A
comparison of the  two correlations  is performed in the present Note.

\section{Correlation of the drag coefficient and  effective \\ Reynolds 
number for the
turbulent boundary layer}

According to the model presented by us previously (see (3)) the drag 
coefficient for
the turbulent pipe flows $c$ as a function of the Reynolds number $Re={\bar u}
d/\nu$ is represented in the form
$$
c =\frac{8u_*^2}{{\bar u}^2} =\frac{8}{\Psi^{2/(1+\a)}} \ , \qquad
\a = \frac{3}{2\ln Re} \ .
$$
Here ${\bar u}$ is the mean velocity (bulk discharge per unit time divided
by the cross-section area), and
$$
\Psi(\a) \ = \ \frac{e^{3/2}(\sqrt{3}+5\a)}{2^\a \a(1+\a)(2+\a)} \ .
$$
As $Re\to\i$, \ $\Psi(\a)\simeq (2e^{3/2}/\sqrt{3})\ln Re$, so that 
asymptotically
the relation is valid
$$
c\simeq\frac{6}{e^3\ln^2 Re}\simeq\frac{0.3}{\ln^2Re} \ . \eqno[3]
$$

The relation [3] is asymptotically covariant. This means that a 
replacement of $Re$
by  certain $Re'$, equal to Constant times $Re$, leaves the 
asymptotics [3] invariant.
Contrary to that, a redefinition of $c$, e.g.~an introduction of a 
different factor
before $u_*^2/{\bar u}^2$ or the use of the maximum velocity instead 
of mean velocity
${\bar u}$, will change the Reynolds-number dependence of the drag coefficient.

Note that at large Reynolds number the expression for the drag 
coefficient should be
identical for all shear flows provided the  Reynolds number is 
properly determined.
This identity was demonstrated for velocity profiles (1,2). Therefore it is
logical to suggest that the expression for the drag coefficient in 
the boundary layer
be of the form
$$
c\,'_f =\frac{2u_*^2}{U^2} = \frac{\mb{Const}}{\ln^2 Re} \ . \eqno[4]
$$
Here $Re$ is the effective Reynolds-number for the boundary layer, 
introduced in
(1,2).

Processing of the experimental data by Naguib (6) and Nagib and Hites (see (7))
confirms (see Figure 1) the correlation [4], and gives the value of 
the Constant in
[4] equal to 0.26, so that the correlation [4] takes the form
$$
c\,'_f =\frac{0.26}{(\ln Re)^2} \ . \eqno[5]
$$

At the same time the data of \"Osterlund (8) reveal a systematic deviation from
[5]\linebreak (Figure 2). We suggest one of the reasons for this 
disagreement is the
following fact: The friction velocity $u_*$ was measured in (6,7) 
whereas it was
calculated in (8) using the universal logarithmic law with 
inappropriately low values
of the constants.

\section{The correlation between the drag coefficient $c\,'_f$ and the
Reynolds number $Re_{\th}$ based on the momentum thickness}

Watson, Hall and Anders (5) proposed recently the following 
correlation between the
drag coefficient  $c\,'_f$ and the traditional  Reynolds
number $Re_{\th}$:
$$
c\,'_f = 0.0097  Re_{\th}^{-0.144} \ . \eqno[6]
$$
The correlation [6] was based on different experimental data covering 
the range of
$Re_{\th}$ from $3\dt 10^4$ to $6\dt 10^5$.

Our comparison showed (see Figures 3 and 4) that both the data of the experimentsof (6),(7) (Illinois Institute of Technology) and (8) (Roya 
l Institute of Technology,
Stockholm) disagree systematically with the Langley correlation [6].
The comparison  of our correlation [5] with the results of Langley 
experiments was
for us impossible due to the incompleteness of the data available to us.


We notice that the power type asymptotics with small values of the 
exponent similar
to [6] can be approximated by  formulas similar to [5] and vice versa.
Indeed, let us approximate $1/\ln^2x$ by a power function
$$
\frac{1}{\ln^2x}=Gx^{-\g} \ .
$$
The constants $G$ and $\g$ can be determined for instance from the 
condition that at
a certain point $x_0$ the functions $Gx^{-\g}$ and $1/\ln^2x$ be 
equal as well as
their derivatives. This condition gives $\g=2/\ln x_0$, and
$G=x_0^\g(\ln x_0)^{-2}$. So, if the correlation [5] has a physical 
meaning, the
empirically obtained correlations of the power type similar to [6] 
can be in fact
their approximations.

\section{Conclusion}

A correlation between the drag
coefficient and the effective Reynolds number for the turbulent boundary layers
at large Reynolds numbers is proposed. It is in satisfactory 
agreement with the data
of Naguib (6), and Nagib and Hites (7).

The data of \"Osterlund (8) for the drag coefficient as a function of 
the effective
Reynolds number are in systematic disagreement with the correlation [5].
Therefore there is a systematic disagreement between the data (6),(7) 
on one hand,
and (8) on the other.

All the data of (6), (7) and (8) are is substantial systematic 
disagreement with the
correlation [6] proposed in (5).  A detailed analysis of the 
experimental data of (5)
is needed. If such a correlation can be firmly established, it will provide the
relation between the characteristic length scale $\Lambda$ and the 
momentum thickness
$\th$.

\bs{\bf Acknowledgments.} This work was supported in part by the
Applied Mathematics subprogram of the U.S.~Department of Energy under
contract DE--AC03--76--SF00098, and in part by the National Science
Foundation under grants DMS\,94--16431 and DMS\,97--32710.

\bs\begin{center}{\bf References}\end{center}

\begin{enumerate}\item Barenblatt, G.I., Chorin, A.J., Hald, O.~and
Prostokishin, V.M. (1997). {\it Proc. Nat. Acad. Sci. USA} {\bf 94}, 
7817--7819.
\item  Barenblatt, G.I., Chorin, A.J.~and
Prostokishin, V.M. (2000). {\it J. Fluid Mech.} {\bf 410}, 263--283.
\item  Barenblatt, G.I., Chorin, A.J.~and
Prostokishin, V.M. (1997). {\it Appl. Mech. Rev.} {\bf 50}, 413--429.
\item  Barenblatt, G.I., Chorin, A.J.~and
Prostokishin, V.M. (2000). {\it Proc. Nat. Acad. Sci. USA} {\bf 97}, 
3799--3802.
\item Watson, R.D., Hall, R.M.~and Anders, J.B. (2000). Review of Skin Friction
Measurements Including Recent High-Reynolds Number Resutls from {\sc 
nasa} Langley
NTF. Paper AIAA--20000-2392, 1--20.
\item Naguib, A.M. (1992). Inner- and outer-layer effects on the dynamics of a
turbulent boundary layer, Ph.D.~Thesis, Illinois Inst.~of Technology, 
Chicago, Ill.
\item Hites, M.H. (1997). Scaling of High-Reynolds Number Turbulent 
Boundary Layers
in the National Diagnostic Facility. Ph.D. Thesis, Illinois Inst.~of 
Technology,
Chicago, Ill.
\item \"Osterlund, J.M. (1999). Experimental studies of zero pressure-gradient
turbulent boundary layer flow, Ph.D.~Thesis, Royal Inst.~of 
Technology, Stockholm,
Sweden.
\end{enumerate}

\newpage

\bs\bs\bs\begin{quote}{\bf Figure Captions}\medskip\\
Figure 1: The correlation [5] is in general agreement with the experimental
data of Naguib (6) (diamonds) and Nagib and Hites (7) (squares).

\bs Figure 2: There is a systematic substantial disagreement of the 
correlation [5]
and the data of \"Osterlund (8) (circles).

\bs Figure 3: There is a systematic disagreement between the correlation [6]
and the experimental data (6,7).

\bs Figure 4: There is a systematic disagreement between the correlation [6]
and the experimental data (8).
\end{quote}

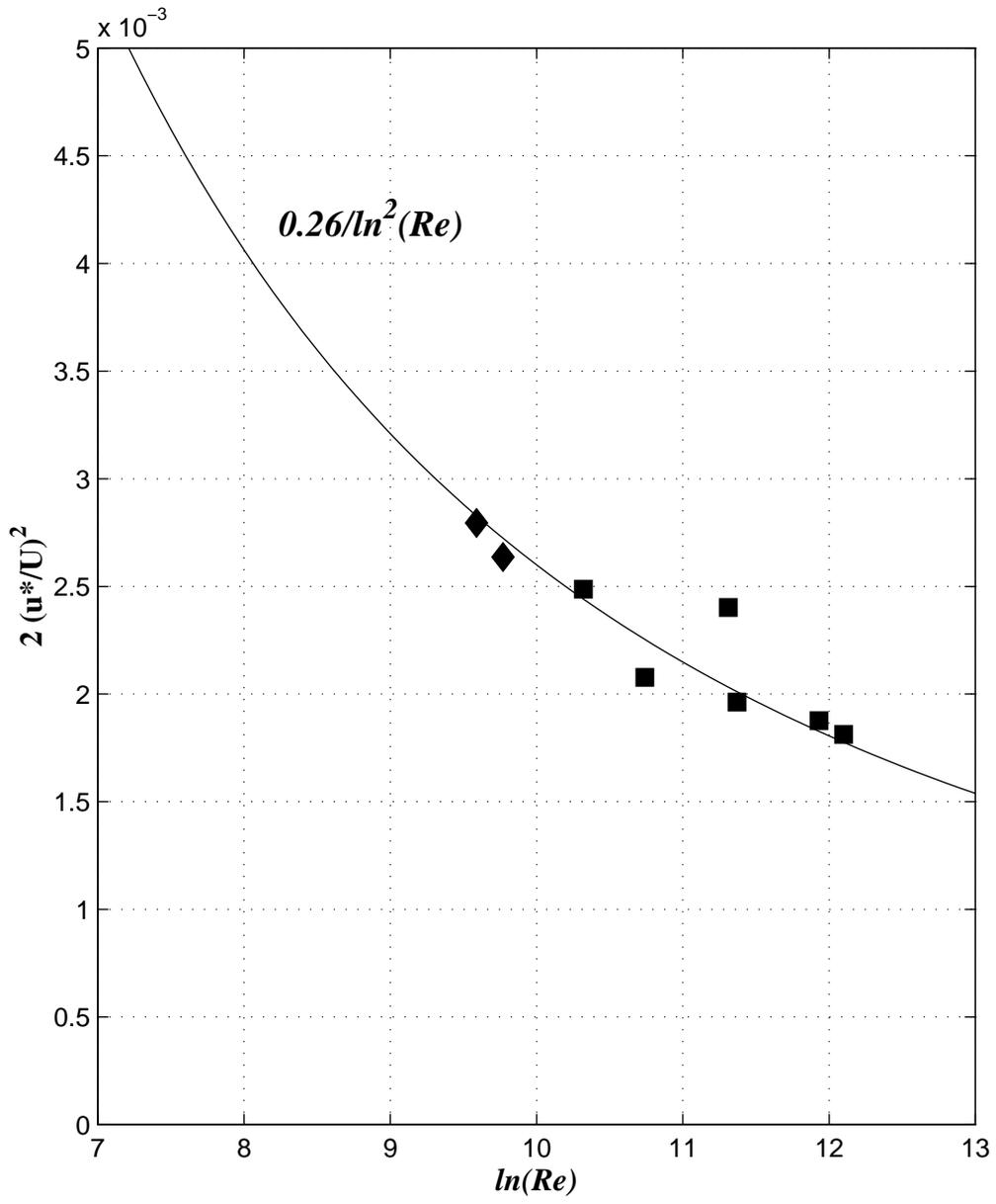
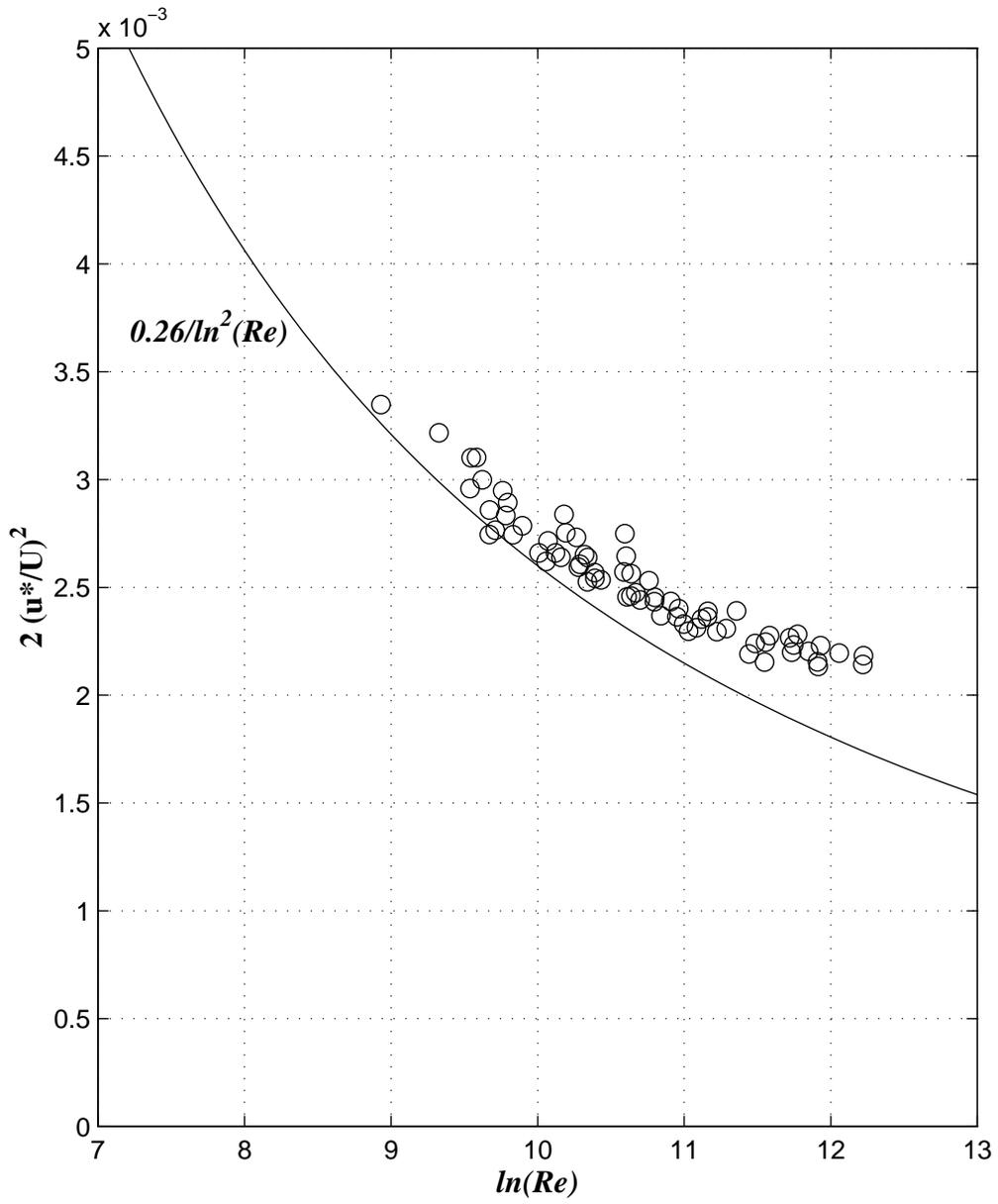
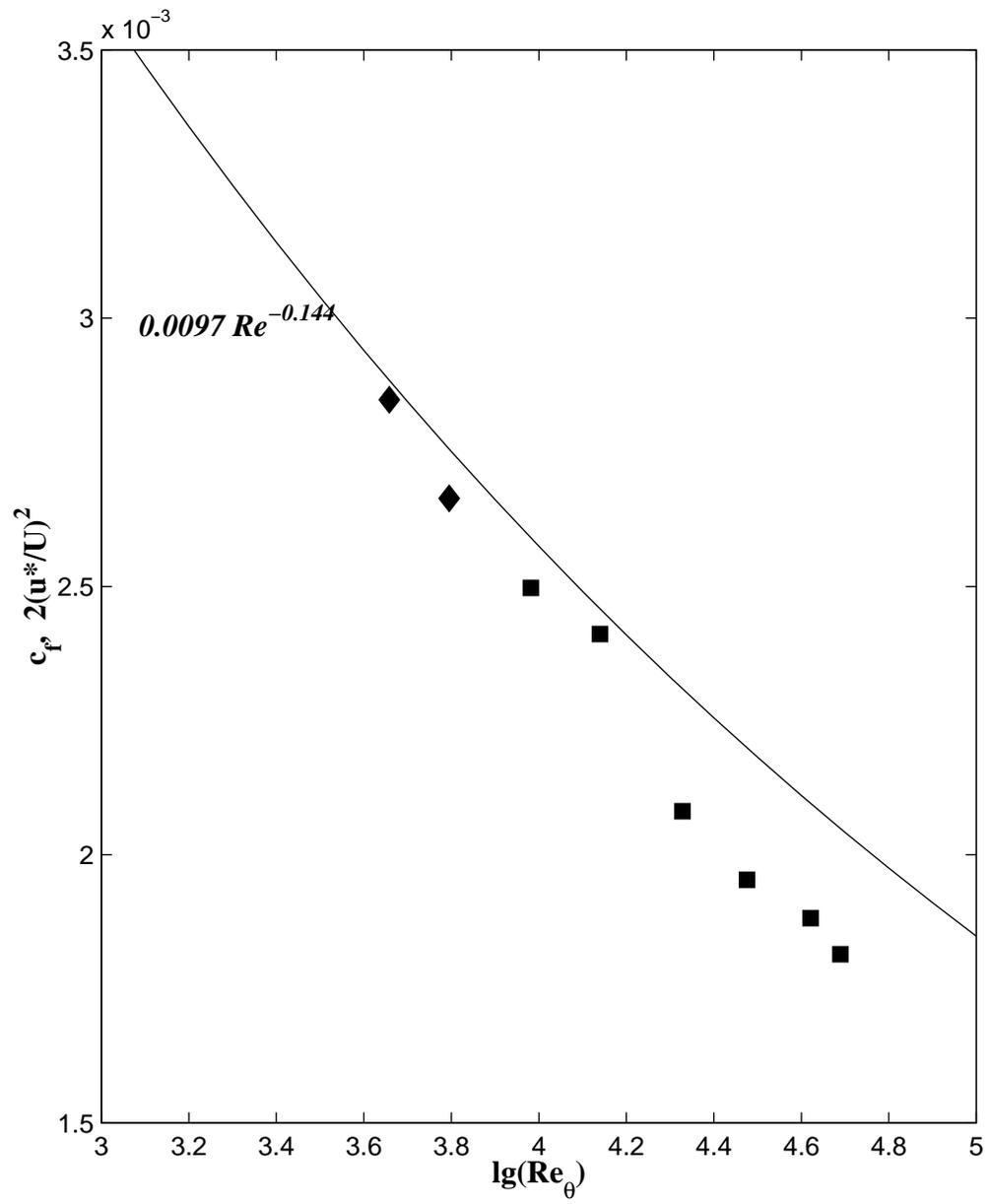
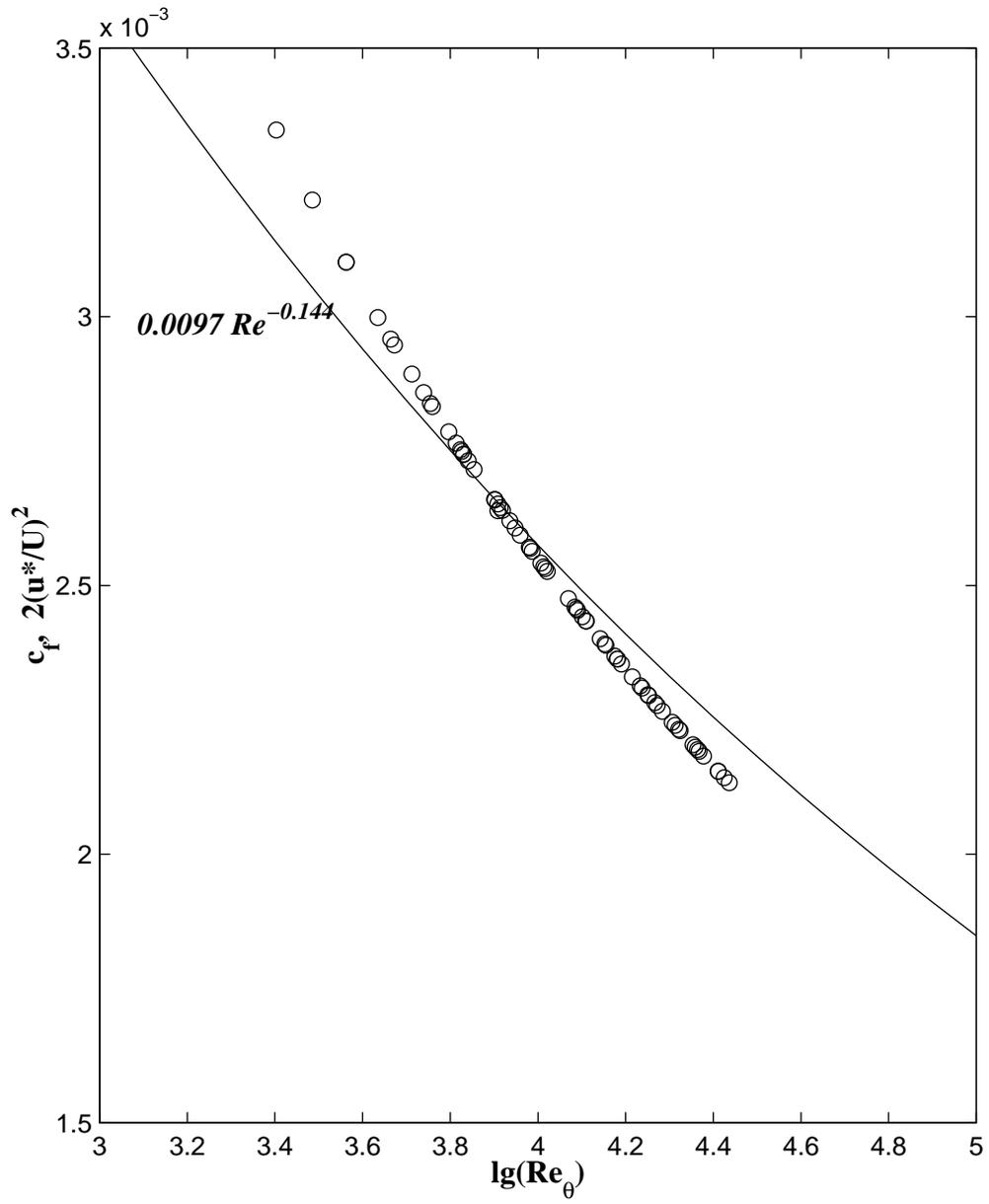
\end{document}